\documentclass[12pt]{article}
\usepackage[dvips]{graphicx}
\usepackage{pdproc}

  \makeatletter 
  \def\@cite#1{[#1]} 
  \makeatother    
\begin{document}

\renewcommand{\thefootnote}{\alph{footnote}}

\title{
Precision Electroweak Physics and Higgsless Models in Warped Space
}

\author{GIACOMO CACCIAPAGLIA}

\address{ 
Inst. for High--Energy Phenomenology\\
Newman Laboratory of Elementary Particle Physics\\
Cornell University, Ithaca, NY 14850, USA
\\ {\rm E-mail: cacciapa@lepp.cornell.edu}}

\abstract{
We study tree level corrections to precision electroweak physics in the recently proposed Higgsless models in warped space.
Such models inherit from their similarity with technicolor theories a large contribution to the oblique parameters, $S$ in particular.
We show that it is possible to suppress $S$ using brane induced kinetic terms and unequal left-right bulk gauge couplings, paying the price of heavy KK modes.
In the allowed region, they are eventually ineffective in restoring perturbative unitarity in $W$ scattering above $2$~TeV.
Although it looks like a Higgsless models' bane, it has been recently shown that such problem can be easily solved delocalizing the light fermions.
}

\normalsize\baselineskip=15pt

\section{Introduction to the Model}

Notwithstanding the amazing success of the Standard Model (SM) in describing high energy physics, we are still missing experimental information about its main ingredient: the mechanism of electroweak symmetry breaking.
This lack has left open space for theoretical speculations and for pursuing more or less radical alternatives.
The main theoretical motivation is the need to stabilize the Higgs mass against radiative correction.
A recent new proposal is the Higgsless scenario~\cite{CGPT}.
In extra dimensions, it is indeed possible to break gauge symmetries via boundary conditions, without any light scalar appearing in the theory.
Now, the scattering amplitude of longitudinal $W$ bosons is unitarized by the gauge boson resonances rather than by the Higgs field~\cite{CGMPT}.
Moreover, the enlarged bulk gauge symmetry $SU(2)_L\times SU(2)_R\times U(1)_{B-L}$  together with a warped background yields a double advantage~\cite{ADMS}: a custodial symmetry protects the correct $M_Z/M_W$ ratio and the warping raises the resonance masses to a realistic level.
Similarly, fermion masses can be generated by boundary conditions~\cite{CGHST}.

Such models also show several similarities with technicolor models via the AdS/CFT correspondence, in particular large oblique corrections are expected.
Indeed, in the simplest model $S$ turns out to be of order one, resulting from the tree level mixing with the KK modes.
Before discussing the details of precision physics, we will briefly summarize the structure of the model~\cite{CGPT,Nomura}.
We will consider a bulk SU(2)$_L\times$SU(2)$_R\times$U(1)$_{B-L}$  
gauge theory
on an AdS$_5$ background, working in the conformally flat metric. 
The AdS curvature $R$ is assumed to be of order $1/M_{Pl}$, however it is a freely  adjustable parameter.
The parameter $R'$ sets the scale of the
gauge boson masses, and will therefore be $R' \sim 1/$TeV. 
We will use the usual bulk Lagrangian, with canonically normalized  
kinetic terms and in the unitary gauge, where all the $A_5$'s decouple  
and we are left with a KK tower of vector fields,  
$(A^{L}_{\mu},A^{R}_{\mu},B_{\mu})$.
We denote the 5D gauge couplings by $g_{5L}$, $g_{5R}$ and  
$\tilde{g}_5$.
Electroweak symmetry breaking is achieved by the boundary conditions  
that break $SU(2)_L \times SU(2)_R \rightarrow SU(2)_D$ on the TeV  
brane and $SU(2)_R \times U(1)_{B-L} \rightarrow U(1)_Y$ on the Planck  
brane.
We also consider kinetic terms allowed on the  
branes, that in terms of field stress tensors can be  
parametrized:
\begin{equation} \label{genLagr}
\mathcal{L} = - \left[ \frac{r}{4} {W^L_{\mu \nu}}^2 +
\frac{r'}{4} {B^Y_{\mu \nu}}^2\right] \delta (z-R)
  - \frac{R'}{R} \left[ \frac{\tau'}{4} {B_{\mu\nu}}^2
+ \frac{\tau}{4} {W^D_{\mu \nu}}^2 \right]\delta (z-R')\,,
\end{equation}
where $A^D = (g_{5R} A^R + g_{5L} A^L)/\sqrt{g_{5R}^2+g_{5L}^2}$ and  
$B^Y = (g_{5R} A^{R3} + \tilde g_5 B)/\sqrt{g_{5R}^2+ \tilde g_{5}^2}$.

One combination of parameters is fixed by the $W$ mass, while the matching of the
4D couplings $g$, $g'$ determines two more parameters. 
Therefore one can
pick as free parameters of the theory the following set: $R$,
$g_{5R}/g_{5L}$, $r$, $r'$, $\tau$, $\tau'$.

\section{Oblique Corrections}

In order to compare Higgsless models to precision electroweak measurements, we need to compute the Peskin-Takeuchi parameters $S$, $T$ and $U$.
We use such parameters to fit the Z-pole observables at LEP1.
In~\cite{BPRS}, Barbieri {\it et al.} proposed an enlarged set of parameters, to take into account also differential cross section measurements at LEP2.
However, the only new information contained by the new parameters is the bound on four-fermi operators generated by the exchange of KK bosons, that we take into account to bound the lighter resonances at LEP2 and Tevatron.
Effectively, our $S$, $T$ and $U$ are linear combinations of the parameters in ~\cite{BPRS}.

In~\cite{CCGT} we computed the oblique corrections in the standard way~\cite{CET}, in terms of mass eigenstates, in the limit where the light fermions are localized on the Planck brane.
The only relevant tecnical point in the calculation, is the matching of the 4D gauge couplings.
Indeed, if one writes down the couplings of the fermions, only two quantities does not depend on the overall $Z$ and $W$ normalizations and are completely fixed by the boundary condition.
Namely, the electric charge and the ratio between the hypercharge and $T_3$ couplings to the $Z$.
Matching such quantities with the SM prediction, it is possible to cast all the corrections in the oblique parameters.

In the basic model, with $g_{5L} = g_{5R} = g_5$ and vanishing localized kinetic terms, the leading contribution to $S$ in the $1/\log \frac{R'}{R} \approx .3$ expansion is:

\begin{equation}
S\approx \frac{ 6 \pi }{g^2\log \frac{R'}{R}}\approx 1.15\,~,
\label{eq:S}
\end{equation}
while $T\approx U \approx 0$.
This value of $S$ is clearly too large to be compared with the experimental result\footnote{Actually, this number should not be compared with the usual SM fit, but we should disentangle the contribution of the Higgs. Namely, it is enough to do the fit assuming a large Higgs mass, equal to the cut-off of the theory~\cite{BPRS}. We are also neglecting loop corrections from the gauge KK modes.}.

As we already mentioned, however, the theory has more parameters.
We first study the effect of asymmetric bulk gauge couplings and Planck brane kinetic terms.
The leading contribution to $S$ is:

\begin{equation}
S\approx \frac{ 6 \pi }{g^2\log \frac{R'}{R}} \frac{2}{1 +
\frac{g_{5R}^2}{g_{5L}^2}} \frac{1}{1+\frac{r}{R \log R'/R}}~,
\label{eq:SPl}
\end{equation}
where, again, $T\approx U \approx 0$.
Now, in case of large $g_{5R}/g_{5L}$ ratio (or large $SU(2)_L$ kinetic term) $S$ is suppressed.
However, the $W$ mass squared is also parametrically multiplied by the same factor.
This means that the smaller $S$ the larger the scale of the KK resonances, $1/R'$.
So, in order to have small corrections we possibly enter a strong coupling regime, where the above calculation becomes meaningless.

Another set of parameters are the TeV kinetic terms.
Their contribution is more complicated, so we will show some results at leading order for $\tau, \tau' \ll R \log \frac{R'}{R}$.
The $SU(2)_D$ kinetic term appears at linear order, and effectively multiplies eq.~\ref{eq:SPl} by a factor $1+\frac{\tau}{R}$.
On the other hand, the $U(1)_{B-L}$ kinetic term contributes at quadratic order.
If only $\tau'$ is turned on,

\begin{eqnarray}
S&\approx&\frac{ 6 \pi }{g^2\log \frac{R'}{R}}-\frac{ 8 \pi }{g^2} 
\left( 1-\left(\frac{g'}{g}\right)^2 \right) \frac{\tau'^2}{(R \log 
R'/R)^2}~, \label{eq:Staup}\\
T&\approx&-\frac{ 2 \pi }{g^2} \left( 1-\left(\frac{g'}{g}\right)^4 
\right) \frac{\tau'^2}{(R \log R'/R)^2}~,
\end{eqnarray}
while $U\approx 0$.
So, $S$ vanishes for $\tau' \approx 0.15\, R \log \frac{R'}{R}$.
However, another effect is to make one of the $Z$' lighter, namely the one that couples with the hypercharge.

\begin{figure}[htb]
\begin{center}
\includegraphics*[width=10cm]{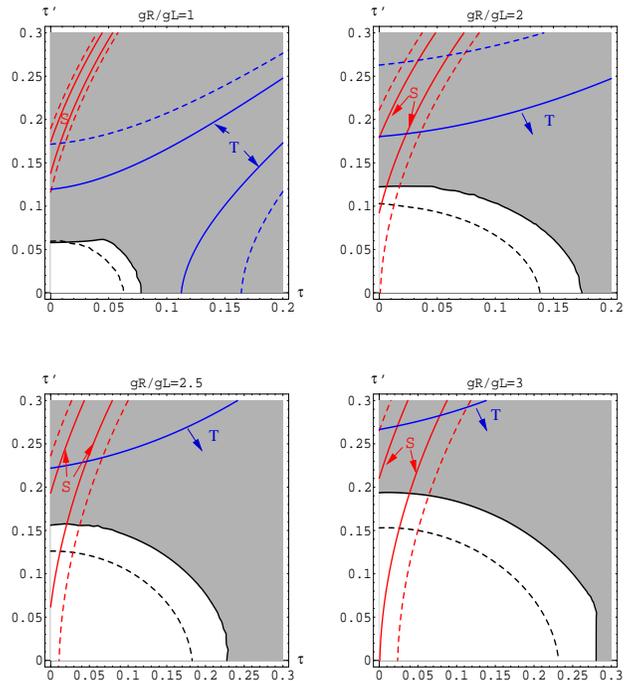}
\caption{%
Contourlines for $|S|$ (red) and $|T|$ (blue) at $0.3$ and $0.5$. The shaded region is excluded by LEP2, allowing a $3$\% deviation in the cross section (the dashed line corresponds to $2$\% deviation).}
\label{fig1}
\end{center}
\end{figure}

We also numerically scanned the parameter space to seek for a region where the model is not ruled out.
For different values of $g_{5R}/g_{5L}$\footnote{Using the Planck kinetic terms instead would only result in slightly different $Z$' couplings, and so different exclusion plots.}, we scanned the $\tau - \tau'$ space (see fig.~\ref{fig1}).
Requiring both $|S|$ and $|T|$ to be smaller that $0.3$, there is an allowed region only for large ratio, $g_{5R}/g_{5L}>2.5$, where the theory is most likely strongly coupled.
These results are in agreement with similar studies in \cite{DHLR2} and \cite{BPRS}.

\section{Conclusions and Outlook}

We have studied the feasibility of the Higgsless models when facing precision electroweak tests.
As originally proposed, the model seems to be disfavoured by the experiments, if one wants strong coupling to arise above $3$~TeV.
However, there is a simple solution that avoids such problems~\cite{CCGT2}, namely to allow the light fermions leaking into the bulk.
In the case of almost flat fermions, $S$ vanishes and the resonances almost decouple with the light fermions.
The direct bounds are then easily avoided and the KK masses can be lowered increasing $R$, thus raising the cut-off of the theory.

\small 
\section{Acknowledgments}

We thank the organizers of SUSY2004 for the support, and Csaba Csaki, Christophe Grojean and John Terning for the collaboration this work is based upon.
This research is supported in part by the DOE OJI grant DE-FG02-01ER41206 and in part by the NSF grants PHY-0139738  and PHY-0098631.

\bibliographystyle{plain}

\end{document}